\documentclass[aip,apl,reprint]{revtex4-1}
\usepackage{graphicx}
\usepackage{dcolumn}
\usepackage{bm}

\draft 

\begin{document}

\title{A three step laser stabilization scheme for excitation to Rydberg levels in $^{85}$Rb} 

\author{L A M Johnson}
\author{H O Majeed}
\author{B T H Varcoe}
\affiliation{School of Physics and Astronomy, University of Leeds, Leeds, LS2 9JT, UK}

\date{\today}

\begin{abstract}


We demonstrate a three step laser stabilization scheme for excitation to $n$P and $n$F Rydberg states in $^{85}$Rb, with all three lasers stabilized using active feedback to independent Rb vapor cells. The setup allows stabilization to the Rydberg states 36P$_{3/2}$\,-\,70P$_{3/2}$ and 33F$_{7/2}$\,-\,90F$_{7/2}$, with the only limiting factor being the available third step laser power. We study the scheme by monitoring the three laser frequencies simultaneously against a self-referenced optical frequency comb. The third step laser, locked to the Rydberg transition, displays an Allan deviation of 30\,kHz over 1 second and $<$80\,kHz over 1 hour. The scheme is very robust and affordable, and it would be ideal for carrying out a range of quantum information experiments.    

\end{abstract}


\maketitle 



\section{Introduction}\label{intro}

Long term laser stabilization to atomic levels is useful for making precision measurements in spectroscopy \cite{ye1996,barwood1991} and, for the case of Rydberg levels, is critical for modern Rydberg atom cavity QED proposals \cite{jones2009,blythe2006}. In both cases long term frequency stability is of importance to maintain a high frequency accuracy. It has recently been demonstrated that the detection of Rydberg states in a thermal vapor cell can offer greatly reduced sensitivity to stray electric fields \cite{Mohapatra2008,thoumany2009two,me2010}. Not only this but purely optical detection makes these experiments much simpler to operate than conventional field ionization type experiments which use a beam or trapping apparatus. Another clear advantage is the simplicity of acquiring analogue spectroscopic Rydberg signals for laser stabilization applications, by using an ordinary photo diode.

Recent work has demonstrated laser stabilization to these types of Rydberg signal, detected in Rb vapor cells, using different types of excitation scheme \cite{thoumany2009,abel2009}. In Ref.~\onlinecite{thoumany2009} a V scheme is exploited, where a 297\,nm frequency doubled dye laser is used to excite the UV transition directly from the ground state to Rydberg level, whilst a 780\,nm laser is used for detection on a strong D1 cycling transition, using an electron shelving scheme. This single step laser excitation allows excitation to $n$P Rydberg states. In Ref.~\onlinecite{abel2009} a two step ladder scheme is used, where a 780\,nm laser excites a D1 transition and acts as a weak probe in an EIT system. A 480\,nm frequency doubled diode laser then acts as a strong pump beam on the 5P$_{3/2}$ to Rydberg level transition. This two step laser excitation allows EIT signals to be detected for $n$S and $n$D Rydberg states. In this work we excite Rydberg states in $^{85}$Rb using a three step laser excitation scheme, the detection method is discussed in previous work \cite{me2010,thoumany2009two}. The three step level system consists of a 780\,nm transition 5S$_{1/2}$ $F=3$ to 5P$_{3/2}$ $F=4$, a 776\,nm transition 5P$_{3/2}$ $F=4$ to 5D$_{5/2}$ $F=5$ and finally a 1260\,nm transition 5D$_{5/2}$ to $n$L$_{J}$. Rydberg excitations are detected via the reduced absorption of the first step laser on a photo diode. This three photon scheme may be advantageous over the two discussed above in that it uses only conventional IR diode lasers, with no need for second harmonic generation. This makes the lasers more user-friendly and affordable. Also, the addition of a third laser step allows excitation to the $n$F series of Rydberg states. In this work our frequency comb gives us a unique opportunity to assess the behavior of each of the stabilized lasers in the three step scheme simultaneously.

\section{Apparatus}\label{app}

\begin{figure}[h]
\begin{center}
\includegraphics[width=8.3cm]{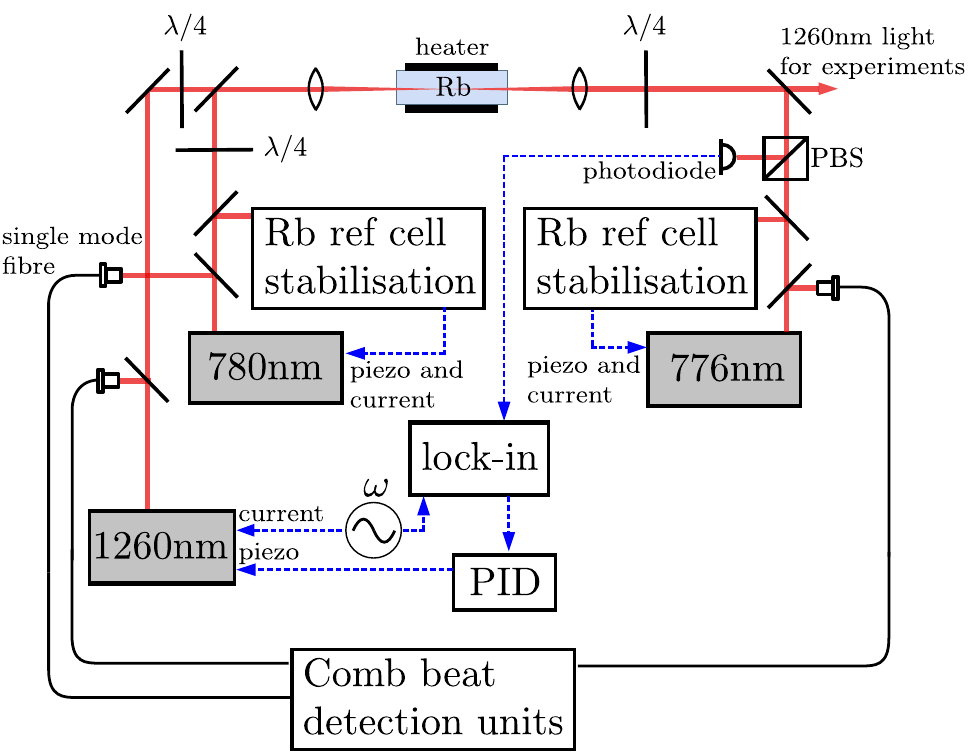}
\caption{The experimental setup used for stabilizing to Rydberg levels. The 780\,nm and 776\,nm lasers are locked to separate Rb reference cells with active feedback to the laser piezo and injection current. The 1260\,nm laser is frequency modulated via the injection current and lock-in detection of the first step absorption is carried out to derive an error signal for frequency stabilization. The 1260\,nm laser is locked using active feedback to the laser piezo. Single mode optical fibers transport laser light from all three lasers to three frequency comb beat detection units.}
\label{Setup}
\end{center}
\end{figure}

Figure \ref{Setup} shows a schematic of the experimental arrangement. The optical setup for Rydberg detection is the same as that described in our previous work \cite{me2010}. One modification has been made in order to use the 1260\,nm laser light in our micromaser experiment after traversing the cell: The introduction and removal of the third step laser light from the setup is achieved using dielectric coated mirrors with polished rear faces. These mirrors reflect only in the 780\,nm region and allow the 3\,mW of 1260\,nm laser power to be removed from the stabilization scheme and sent to our cryostat via an optical fiber. The weak coupling of the third step transition ensures there is negligible attenuation of this laser on passing through the vapor cell.

The first step laser is stabilized using a separate Rb reference cell using a polarization spectroscopy scheme, which allows stabilization to the center of a spectral line without the need for frequency modulation  \cite{pearman2002}. Active feedback for this laser lock is supplied via the laser cavity piezo and the diode injection current.

The second step error signal is derived from the reduced first step absorption, in the same manner as the Rydberg detection. The first and second steps co-propagate through another separate Rb reference cell. When the first step is stabilized it selects only zero velocity atoms from this cell. When the second step is scanned over the 5P$_{3/2}$ $F=4$ to 5D$_{5/2}$ $F=5$ transition we see a Doppler free peak in the first step absorption, corresponding to reduced absorption as atoms are removed from the strong 780\,nm cycling transition. By adding a small frequency modulation to the second step laser, and monitoring the first step absorption via a lock-in amplifier, an error signal is extracted for stabilization to the top of this peak. Feedback for this lock is also supplied via the laser cavity piezo and the diode injection current. 

To derive an error signal for stabilization to the Rydberg levels we add a frequency modulation to the third step laser via the injection current, with a modulation amplitude of 15\,MHz and frequency of 90\,kHz. Detection of the first step absorption is carried out using a lock-in amplifier, with a time constant of 100\,$\mu$s. Active feedback for this lock is supplied via the laser cavity piezo. For all three stabilization schemes, the error signals are sent through Proportional Integral Differential (PID) controllers and then to the laser heads for feedback.  

A proportion of the light from each of the lasers is sent to individual comb beat detection units where the laser light is mixed with light from a self referenced optical frequency comb. The frequency comb spectrum is centered at 1500\,nm. Second harmonic generation and spectral amplification is used to supply comb light at 780\,nm, 776\,nm and 1260\,nm. In each beat detection unit the beat note between the laser and a predetermined comb line is detected with a pin photo diode at around 20\,MHz. The beat notes are subsequently amplified and filtered before being counted with frequency counters. All three frequency counters are synchronized to take readings simultaneously. The counters and stabilization electronics for the frequency comb are all referenced to a Rb frequency standard, disciplined by a GPS receiver. When required, the comb system allows laser frequencies to be measured with an absolute accuracy of $\sim$10$^{-12}$.

\section{Results}\label{results}

As expected, we find that lower $n$ states give Rydberg locking signals with a much larger signal to noise ratio, due to stronger coupling with the third step laser. 
Also, the $n$F$_{7/2}$ signals have a larger amplitude in comparison to the $n$P$_{3/2}$ states for equal laser powers, generally by a factor of two. With this setup it was possible to stabilize to the Rydberg states 36P$_{3/2}$\,-\,70P$_{3/2}$ and 33F$_{7/2}$\,-\,90F$_{7/2}$. For locking to higher $n$ states the only limiting factor was our available third step laser power of 3\,mW. We found that a 15\,MHz modulation amplitude gave an error signal with excellent signal to noise ratio for locking. This modulation would ideally be applied externally with an EOM for example, to leave the 1260\,nm laser linewidth intact for experiments. A simpler, and perhaps more effective option, would be to modulate the atoms using a solenoid around the cell. In this case a dither-free Zeeman lock could be used \cite{weis1988}.

\begin{figure}[!h]
\begin{center}
\includegraphics[width=8.3cm]{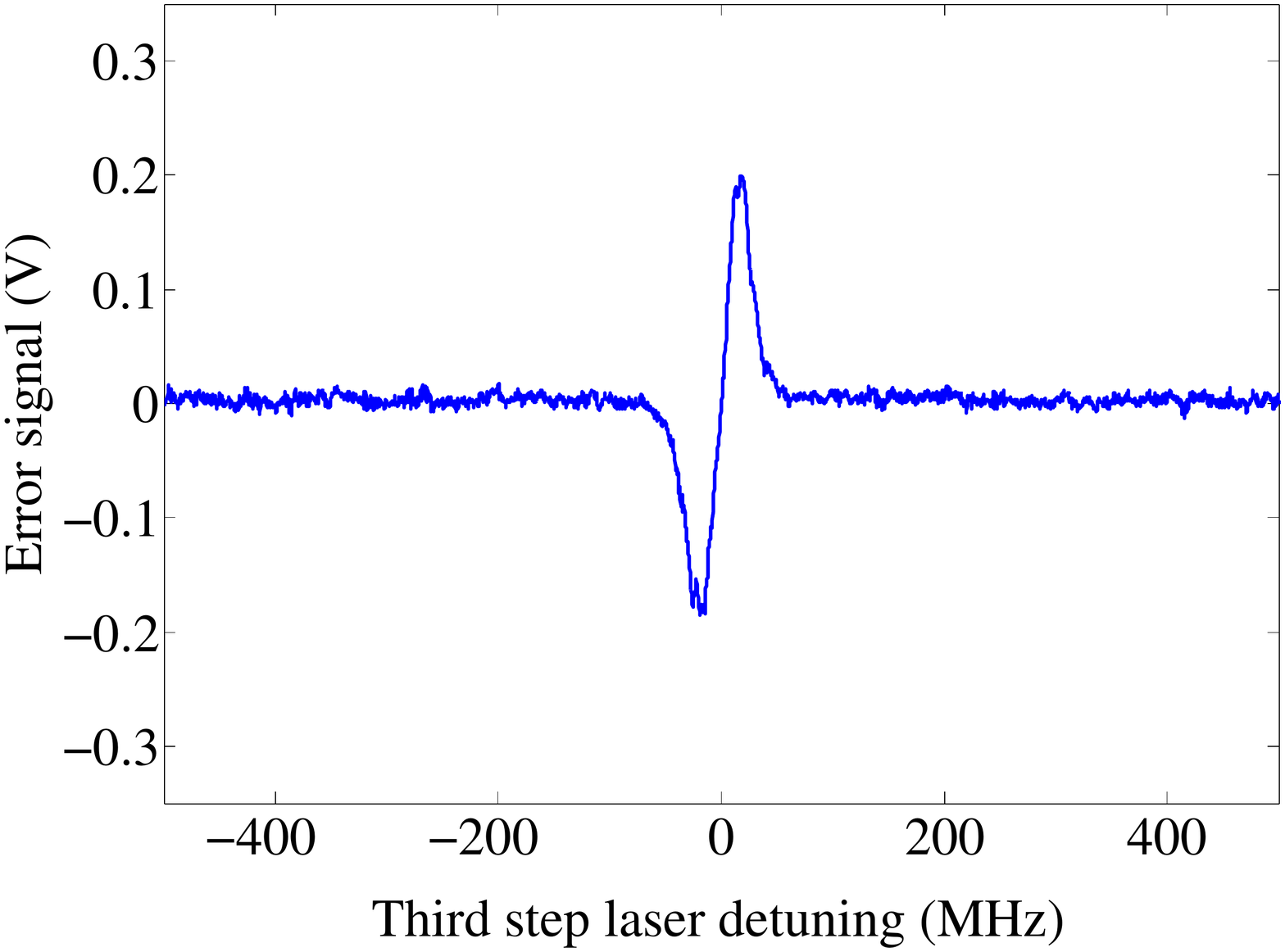}
\caption{The error signal used to stabilize the third step laser to the center of the the 50F$_{7/2}$ transition.}
\label{error}
\end{center}
\end{figure}

The locking signals that were used for locking the third step laser typically had a gradient of 10\,mV/MHz. Figure \ref{error} displays the locking signal used for locking to the 50F$_{7/2}$ Rydberg level. We also chose to study the lock to the 63P$_{3/2}$ level as it forms the lower level of the microwave transition which is used in our micromaser experiment.

\begin{figure}[!h]
\begin{center}
\includegraphics[width=8.3cm]{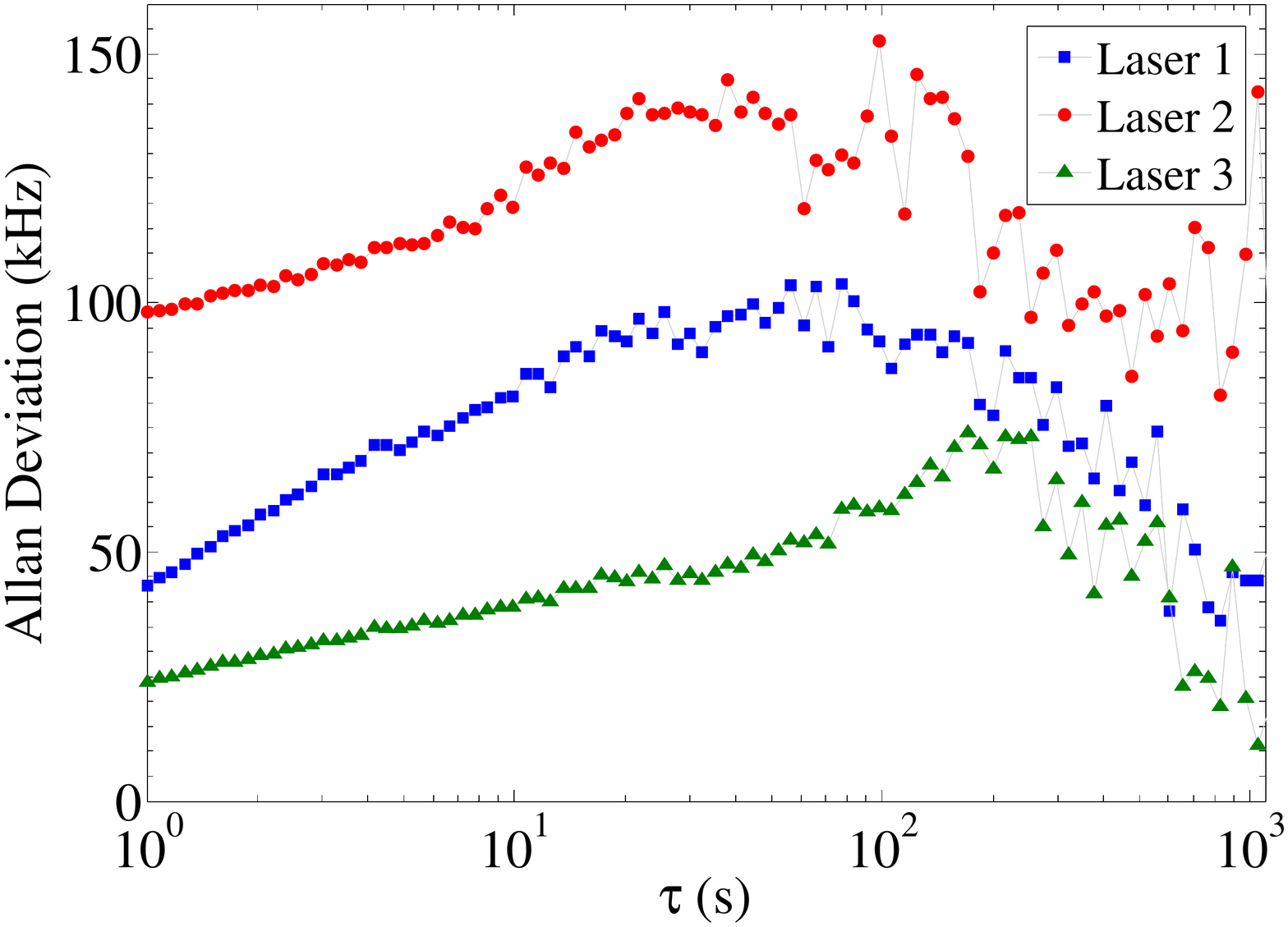}
\caption{The Allan deviation of the three laser locks when locked to their respective cells. This data was computed from a 5000\,s data set. The third step laser was locked to the 50F$_{7/2}$ transition.}
\label{allan}
\end{center}
\end{figure}

We analyzed the stability of the system using the comb counter readings. 
Figure \ref{allan} shows the computed Allan deviations for all three laser steps, from a 5000 second data set, when the third step laser was locked on to the 50F$_{7/2}$ Rydberg level. It can be seen that the Allan deviation of the third step laser is below 80\,kHz for time scales upto 10$^{3}$ seconds, and is below 30\,kHz over one second. We found the Allan deviation of the third step laser when locked to the 63P$_{3/2}$ transition was comparable and stayed below 150\,kHz for time scales upto 10$^{3}$ seconds. These results demonstrate that the Rydberg atoms are very well isolated from the environment, especially as this vapor cell is unshielded from magnetic and electric fields.

As one would expect, we find that frequency fluctuations are correlated between each of the laser steps: The first step laser selects a narrow 6\,MHz velocity range of atoms from the thermal distribution, with which the other two lasers interact. We also used the frequency comb to study the behavior of the system to laser detunings. For the first two steps, which have similar wavelengths, we see an almost one-to-one correlation between the first step detuning and the shift of the second step signal, this occurs due to the close Doppler matching of these two steps, with a wavelength ratio of 1.01. This relationship is not clear from figure \ref{allan}, where the Allan deviation for the second step lock is considerably higher than the first step, across all time scales. This can be associated with different feedback responses for the two locks. Surprisingly we have also observed that the second step detuning from the observed second step signal causes only very small shifts of the observed Rydberg signal frequency by $\sim$0.1$\times$ the detuning. Therefore the second laser plays only a small part in the third laser's stability. For this reason, we see in figure \ref{allan} that the stability of the third step laser is not compromised by that of the poorer second step lock. Finally, figure \ref{allan} also clearly shows that the third step fluctuations are considerably lower than for the first step. This is an advantage of this particular excitation scheme and is a consequence of the wavelength mismatch between these lasers. Detuning the first step laser causes a shift in the velocity selection from zero, this velocity has a Doppler shift which is a factor of 0.6$\times$ smaller at 1260\,nm. Therefore fluctuations of the first step laser are suppressed by a factor of 0.6$\times$ on the third.

Interestingly, if these measurements were translated to absolute optical frequencies they would far surpass the accuracy of those from the best optical Rydberg spectroscopy\cite{me2010,sanguinetti2009}, and would give an accuracy comparable to relative millimeter wave measurements \cite{han2006}. In this study we were unable to be confident of the absolute accuracy of the second and third step counter readings due to observed counting errors of upto 1\,MHz for these frequency modulated beat notes, see for example Ref.~\onlinecite{lee2007}. Future improvements in the setup should be able to overcome this.      

\section{Conclusion}\label{conc}

We have used a frequency comb to demonstrate the stability of a three step laser excitation scheme for excitation to $n$P and $n$F Rydberg states. The results show that the third step laser in the ladder scheme can have a much higher frequency stability than the first step due to the mismatch in wavelengths. The results also show that the second laser lock does not compromise the stability of the third in this scheme. This locking scheme will be of great use in experiments that require $n$P and $n$F Rydberg states. The setup is very easy to construct and maintain due to the fact that all lasers are ordinary diode lasers, with no need for frequency doubling. Our results show that excellent stabilities are achievable with these cell based Rydberg state locks, in this scheme we were able to lock on to the line center of a Rydberg $n$F state, with deviations of $<$80\,kHz over an hour time period.


\begin{thebibliography}{13}%
\makeatletter
\providecommand \@ifxundefined [1]{%
 \@ifx{#1\undefined}
}%
\providecommand \@ifnum [1]{%
 \ifnum #1\expandafter \@firstoftwo
 \else \expandafter \@secondoftwo
 \fi
}%
\providecommand \@ifx [1]{%
 \ifx #1\expandafter \@firstoftwo
 \else \expandafter \@secondoftwo
 \fi
}%
\providecommand \natexlab [1]{#1}%
\providecommand \enquote  [1]{``#1''}%
\providecommand \bibnamefont  [1]{#1}%
\providecommand \bibfnamefont [1]{#1}%
\providecommand \citenamefont [1]{#1}%
\providecommand \href@noop [0]{\@secondoftwo}%
\providecommand \href [0]{\begingroup \@sanitize@url \@href}%
\providecommand \@href[1]{\@@startlink{#1}\@@href}%
\providecommand \@@href[1]{\endgroup#1\@@endlink}%
\providecommand \@sanitize@url [0]{\catcode `\\12\catcode `\$12\catcode
  `\&12\catcode `\#12\catcode `\^12\catcode `\_12\catcode `\%12\relax}%
\providecommand \@@startlink[1]{}%
\providecommand \@@endlink[0]{}%
\providecommand \url  [0]{\begingroup\@sanitize@url \@url }%
\providecommand \@url [1]{\endgroup\@href {#1}{\urlprefix }}%
\providecommand \urlprefix  [0]{URL }%
\providecommand \Eprint [0]{\href }%
\providecommand \doibase [0]{http://dx.doi.org/}%
\providecommand \selectlanguage [0]{\@gobble}%
\providecommand \bibinfo  [0]{\@secondoftwo}%
\providecommand \bibfield  [0]{\@secondoftwo}%
\providecommand \translation [1]{[#1]}%
\providecommand \BibitemOpen [0]{}%
\providecommand \bibitemStop [0]{}%
\providecommand \bibitemNoStop [0]{.\EOS\space}%
\providecommand \EOS [0]{\spacefactor3000\relax}%
\providecommand \BibitemShut  [1]{\csname bibitem#1\endcsname}%
\let\auto@bib@innerbib\@empty
\bibitem [{\citenamefont {{Ye, Jun}}\ \emph {et~al.}(1996)\citenamefont {{Ye,
  Jun}}, \citenamefont {{Swartz, Steve}}, \citenamefont {{Jungner, Peter}},\
  and\ \citenamefont {{Hall, John L.}}}]{ye1996}%
  \BibitemOpen
  \bibfield  {author} {\bibinfo {author} {\bibnamefont {{J. Ye}}}, \bibinfo
  {author} {\bibnamefont {{S. Swartz}}}, \bibinfo {author} {\bibnamefont
  {{P. Jungner}}}, \ and\ \bibinfo {author} {\bibnamefont {{J. L. Hall}}},\ }\href {{http://www.opticsinfobase.org/abstract.cfm?id=45035}}
  {\bibfield  {journal} {\bibinfo  {journal} {{Opt. Lett.}}\ }\textbf {\bibinfo
  {volume} {21}},\ \bibinfo {pages} {1280} (\bibinfo {year}
  {1996})}\BibitemShut {NoStop}%
\bibitem [{\citenamefont {Barwood}, \citenamefont {Gill},\ and\ \citenamefont
  {Rowley}(1991)}]{barwood1991}%
  \BibitemOpen
  \bibfield  {author} {\bibinfo {author} {\bibfnamefont {G.~P.}\ \bibnamefont
  {Barwood}}, \bibinfo {author} {\bibfnamefont {P.}~\bibnamefont {Gill}}, \
  and\ \bibinfo {author} {\bibfnamefont {W.~R.~C.}\ \bibnamefont {Rowley}},\
  }\href {\doibase 10.1007/BF00330229} {\bibfield  {journal} {\bibinfo
  {journal} {Appl. Phys. B}\ }\textbf {\bibinfo {volume}
  {53}},\ \bibinfo {pages} {142} (\bibinfo {year} {1991})}\BibitemShut
  {NoStop}%
\bibitem [{\citenamefont {Jones}, \citenamefont {Wilkes},\ and\ \citenamefont
  {Varcoe}(2009)}]{jones2009}%
  \BibitemOpen
  \bibfield  {author} {\bibinfo {author} {\bibfnamefont {M.~L.}\ \bibnamefont
  {Jones}}, \bibinfo {author} {\bibfnamefont {G.~J.}\ \bibnamefont {Wilkes}}, \
  and\ \bibinfo {author} {\bibfnamefont {B.~T.~H.}\ \bibnamefont {Varcoe}},\
  }\href {\doibase 10.1088/0953-4075/42/14/145501} {\bibfield  {journal}
  {\bibinfo  {journal} {J. Phys. B: At. Mol. Opt. Phys.}\ }\textbf {\bibinfo {volume} {42}},\ \bibinfo {pages} {145501}
  (\bibinfo {year} {2009})}\BibitemShut {NoStop}%
\bibitem [{\citenamefont {Blythe}\ and\ \citenamefont
  {Varcoe}(2006)}]{blythe2006}%
  \BibitemOpen
  \bibfield  {author} {\bibinfo {author} {\bibfnamefont {P.~J.}\ \bibnamefont
  {Blythe}}\ and\ \bibinfo {author} {\bibfnamefont {B.~T.~H.}\ \bibnamefont
  {Varcoe}},\ }\href {\doibase 10.1088/1367-2630/8/10/231} {\bibfield
  {journal} {\bibinfo  {journal} {New. J. Phys.}\ }\textbf {\bibinfo {volume}
  {8}},\ \bibinfo {pages} {231} (\bibinfo {year} {2006})}\BibitemShut
  {NoStop}%
\bibitem [{\citenamefont {Mohapatra}, \citenamefont {Jackson},\ and\
  \citenamefont {Adams}(2007)}]{Mohapatra2008}%
  \BibitemOpen
  \bibfield  {author} {\bibinfo {author} {\bibfnamefont {A.~K.}\ \bibnamefont
  {Mohapatra}}, \bibinfo {author} {\bibfnamefont {T.~R.}\ \bibnamefont
  {Jackson}}, \ and\ \bibinfo {author} {\bibfnamefont {C.~S.}\ \bibnamefont
  {Adams}},\ }\href {\doibase 10.1103/PhysRevLett.98.113003} {\bibfield
  {journal} {\bibinfo  {journal} {Phys. Rev. Lett.}\ }\textbf {\bibinfo
  {volume} {98}},\ \bibinfo {pages} {113003} (\bibinfo {year}
  {2007})}\BibitemShut {NoStop}%
\bibitem [{\citenamefont {Thoumany}\ \emph
  {et~al.}(2009{\natexlab{b}})\citenamefont {Thoumany}, \citenamefont
  {Germann}, \citenamefont {H\"{a}nsch}, \citenamefont {Stania}, \citenamefont
  {Urbonas},\ and\ \citenamefont {Becker}}]{thoumany2009two}%
  \BibitemOpen
  \bibfield  {author} {\bibinfo {author} {\bibfnamefont {P.}~\bibnamefont
  {Thoumany}}, \bibinfo {author} {\bibfnamefont {T.}~\bibnamefont {Germann}},
  \bibinfo {author} {\bibfnamefont {T.}~\bibnamefont {H\"{a}nsch}}, \bibinfo
  {author} {\bibfnamefont {G.}~\bibnamefont {Stania}}, \bibinfo {author}
  {\bibfnamefont {L.}~\bibnamefont {Urbonas}}, \ and\ \bibinfo {author}
  {\bibfnamefont {T.}~\bibnamefont {Becker}},\ }\href {\doibase
  10.1080/09500340903180525} {\bibfield  {journal} {\bibinfo  {journal}
  {J. Mod. Opt.}\ }\textbf {\bibinfo {volume} {56}},\ \bibinfo
  {pages} {2055} (\bibinfo {year} {2009}{\natexlab{b}})}\BibitemShut {NoStop}%
\bibitem [{\citenamefont {{Johnson, L. A. M.}}\ \emph
  {et~al.}(2010)\citenamefont {{Johnson, L. A. M.}}, \citenamefont {{Majeed, H.
  O.}}, \citenamefont {{Sanguinetti, B.}}, \citenamefont {{Becker, Th}},\ and\
  \citenamefont {{Varcoe, B. T. H.}}}]{me2010}%
  \BibitemOpen
  \bibfield  {author} {\bibinfo {author} {\bibnamefont {{L. A. M. Johnson}}},
  \bibinfo {author} {\bibnamefont {{H. O. Majeed}}}, \bibinfo {author}
  {\bibnamefont {{B. Sanguinetti}}}, \bibinfo {author} {\bibnamefont {{Th. Becker}}}, \ and\ \bibinfo {author} {\bibnamefont {{B. T. H. Varcoe}}},\ }\href
  {\doibase 10.1088/1367-2630/12/6/063028} {\bibfield  {journal} {\bibinfo
  {journal} {{New. J. Phys.}}\ }\textbf {\bibinfo {volume} {12}},\
  \bibinfo {pages} {063028} (\bibinfo {year} {2010})}\BibitemShut {NoStop}%
\bibitem [{\citenamefont {Thoumany}\ \emph
  {et~al.}(2009{\natexlab{a}})\citenamefont {Thoumany}, \citenamefont
  {H\"{a}nsch}, \citenamefont {Stania}, \citenamefont {Urbonas},\ and\
  \citenamefont {Becker}}]{thoumany2009}%
  \BibitemOpen
  \bibfield  {author} {\bibinfo {author} {\bibfnamefont {P.}~\bibnamefont
  {Thoumany}}, \bibinfo {author} {\bibfnamefont {T.}~\bibnamefont
  {H\"{a}nsch}}, \bibinfo {author} {\bibfnamefont {G.}~\bibnamefont {Stania}},
  \bibinfo {author} {\bibfnamefont {L.}~\bibnamefont {Urbonas}}, \ and\
  \bibinfo {author} {\bibfnamefont {T.}~\bibnamefont {Becker}},\ }\href
  {\doibase 10.1364/OL.34.001621} {\bibfield  {journal} {\bibinfo  {journal}
  {Opt. Lett.}\ }\textbf {\bibinfo {volume} {34}},\ \bibinfo {pages} {1621}
  (\bibinfo {year} {2009}{\natexlab{a}})}\BibitemShut {NoStop}%
\bibitem [{\citenamefont {{Abel, R. P.}}\ \emph {et~al.}(2009)\citenamefont
  {{R. P. Abel}}, \citenamefont {{A. K. Mohapatra}}, \citenamefont {{M. G. Bason}}, \citenamefont {{J. D. Pritchard}}, \citenamefont {{K. J. Weatherill, }}, \citenamefont {{U. Raitzsch}},\ and\ \citenamefont {{C. S. Adams, }}}]{abel2009}%
  \BibitemOpen
  \bibfield  {author} {\bibinfo {author} {\bibnamefont {{R. P. Abel}}},
  \bibinfo {author} {\bibnamefont {{A. K. Mohapatra}}}, \bibinfo {author}
  {\bibnamefont {{M. G. Bason}}}, \bibinfo {author} {\bibnamefont {{J. D. Pritchard}}}, \bibinfo {author} {\bibnamefont {{K. J. Weatherill}}}, \bibinfo
  {author} {\bibnamefont {{U. Raitzsch}}}, \ and\ \bibinfo {author}
  {\bibnamefont {{C. S. Adams}}},\ }\href {\doibase 10.1063/1.3086305}
  {\bibfield  {journal} {\bibinfo  {journal} {{Appl. Phys. Lett.}}\
  }\textbf {\bibinfo {volume} {94}},\ \bibinfo {pages} {071107} (\bibinfo
  {year} {2009})}\BibitemShut {NoStop}%
\bibitem [{pea(2002)}]{pearman2002}%
  \BibitemOpen
  \bibfield  {author} {\bibinfo {author} {\bibnamefont {{C. P. Pearman}}},
  \bibinfo {author} {\bibnamefont {{C. S. Adams}}}, \bibinfo {author}
  {\bibnamefont {{S. G. Cox}}}, \bibinfo {author} {\bibnamefont {{P. F. Griffin}}}, \bibinfo {author} {\bibnamefont {{D. A. Smith}}}, \ and\ \bibinfo {author}
  {\bibnamefont {{I. G. Hughes}}},\ }
  \href {\doibase 10.1088/0953-4075/35/24/315} {\bibfield  {journal} {\bibinfo
  {journal} {{J. Phys. B: At. Mol. Opt. Phys.}}\
  }\textbf {\bibinfo {volume} {35}},\ \bibinfo {pages} {5141} (\bibinfo {year}
  {2002})}\BibitemShut {NoStop}%
\bibitem [{\citenamefont {{Weis, A.}}\ and\ \citenamefont {{Derler,
  S.}}(1988)}]{weis1988}%
  \BibitemOpen
  \bibfield  {author} {\bibinfo {author} {\bibnamefont {{A. Weis}}},\ and\
  \bibinfo {author} {\bibnamefont {{S. Derler}}},\ }\href {\doibase
  10.1364/AO.27.002662} {\bibfield  {journal} {\bibinfo  {journal} {{Appl.
  Opt.}}\ }\textbf {\bibinfo {volume} {27}},\ \bibinfo {pages} {2662} (\bibinfo
  {year} {1988})}\BibitemShut {NoStop}%
\bibitem [{\citenamefont {{Sanguinetti, B.}}, \citenamefont {{Majeed, H. O.}}, \citenamefont {{Jones, M. L.}},\ and\
  \citenamefont {{Varcoe, B. T. H.}}(2009)}]{sanguinetti2009}%
  \BibitemOpen
  \bibfield  {author} {\bibinfo {author} {\bibnamefont {{B. Sanguinetti}}},
  \bibinfo {author} {\bibnamefont {{H. O. Majeed}}}, \bibinfo {author} {\bibnamefont {{M. L. Jones}}}, \ and\ \bibinfo {author}
  {\bibnamefont {{B. T. H. Varcoe}}},\ }\href {\doibase 10.1364/AO.46.000930}
  {\bibfield  {journal} {\bibinfo  {journal} {{J. Phys. B: At. Mol. Opt. Phys.}}\ }\textbf {\bibinfo
  {volume} {42}},\ \bibinfo {pages} {165004} (\bibinfo {year} {2009})}\BibitemShut
  {NoStop}%
\bibitem [{\citenamefont {Han}\ \emph {et~al.}(2006)\citenamefont {Han},
  \citenamefont {Jamil}, \citenamefont {Norum}, \citenamefont {Tanner},\ and\
  \citenamefont {Gallagher}}]{han2006}%
  \BibitemOpen
  \bibfield  {author} {\bibinfo {author} {\bibfnamefont {J.}~\bibnamefont
  {Han}}, \bibinfo {author} {\bibfnamefont {Y.}~\bibnamefont {Jamil}}, \bibinfo
  {author} {\bibfnamefont {D.~V.~L.}\ \bibnamefont {Norum}}, \bibinfo {author}
  {\bibfnamefont {P.~J.}\ \bibnamefont {Tanner}}, \ and\ \bibinfo {author}
  {\bibfnamefont {T.~F.}\ \bibnamefont {Gallagher}},\ }\href {\doibase
  10.1103/PhysRevA.74.054502} {\bibfield  {journal} {\bibinfo  {journal}
  {Phys. Rev. A.}\ }\textbf {\bibinfo {volume} {74}},\ \bibinfo {pages}
  {054502} (\bibinfo {year} {2006})}\BibitemShut {NoStop}%
\bibitem [{\citenamefont {{Lee, Won-Kyu}}, \citenamefont {{Dae-Su Yee}},\ and\
  \citenamefont {{Ho S. Suh}}(2007)}]{lee2007}%
  \BibitemOpen
  \bibfield  {author} {\bibinfo {author} {\bibnamefont {{Won-Kyu Lee}}},
  \bibinfo {author} {\bibnamefont {{Dae-Su Yee}}}, \ and\ \bibinfo {author}
  {\bibnamefont {{Ho S. Suh}}},\ }\href {\doibase 10.1364/AO.46.000930}
  {\bibfield  {journal} {\bibinfo  {journal} {{Appl. Opt.}}\ }\textbf {\bibinfo
  {volume} {46}},\ \bibinfo {pages} {930} (\bibinfo {year} {2007})}\BibitemShut
  {NoStop}%
\end{thebibliography}
\end{document}